\documentclass[usenames,dvipsnames, 12pt, reqno]{amsart}
\usepackage[foot]{amsaddr}
\usepackage{pdflscape}
\usepackage{afterpage}
\usepackage{rotating}
\usepackage[T1]{fontenc}
\usepackage{ae}
\usepackage{aecompl}
\usepackage{newtxtext}
\usepackage{wrapfig}

\usepackage{amsaddr}
\usepackage[margin=1in]{geometry}
\usepackage{url}
\usepackage{amssymb}
\usepackage{amsthm}
\usepackage[normalem]{ulem}
\usepackage{mathabx}
\usepackage{bbm}
\usepackage{array}
\usepackage[colorlinks=true,urlcolor=NavyBlue,linkcolor=NavyBlue,citecolor=NavyBlue]{hyperref}
\usepackage{amsmath}
\usepackage[inline]{enumitem}

\makeatletter
\def\paragraph{\@startsection{paragraph}{4}%
  \z@\z@{-\fontdimen2\font}%
  {\normalfont\bfseries}}
\makeatother

\usepackage{mathrsfs}
\usepackage{framed}
\usepackage{latexsym}
\usepackage{pb-diagram}
\usepackage[mathscr]{euscript}
\usepackage{graphicx,caption}
\usepackage[all]{xy} 
\usepackage{color}
\usepackage{algorithm}
\usepackage[noend]{algpseudocode}
\usepackage{thmtools,thm-restate}

\newcommand{\hz}{\operatorname{Hz}}
\newcommand{\cm}{\operatorname{cm}}

\newcommand{\Z}{\mathbb{Z}}

\newcommand{\mc}{\mathcal}

\newcommand{\s}{\sigma}

\renewcommand{\t}{\tau}

\newcommand{\defeq}{\overset{\tiny\operatorname{def}}{=}}

\renewcommand{\r}{\rightarrow}

\newcommand{\norm}[1]{\|#1\|} 

\newcommand{\numpf}{\texttt{numPF}}
\newcommand{\sizpf}{\texttt{sizePF}}
\newcommand{\meanf}{\texttt{meanF}}
\newcommand{\dt}{\texttt{dt}}
\newcommand{\firewin}{\texttt{fireWin}}
\newcommand{\fireth}{\texttt{fireTh}}
\newcommand{\poisson}{\texttt{Poisson}}
\newcommand{\binary}{\texttt{Binary}}
\newcommand{\fuzzyb}{\texttt{FuzzyB}}
\newcommand{\longbars}{\texttt{LongBars}}

\theoremstyle{definition}

\title[The Importance of Forgetting: A Topological Perspective]{The Importance of Forgetting: Limiting Memory Improves Recovery of Topological Characteristics from Neural Data}
\author{Samir Chowdhury, Bowen Dai, and Facundo M\'{e}moli}

\begin{document}

\email{chowdhury.57@osu.edu, bowen.dai.gr@dartmouth.edu, memoli@math.osu.edu}

\address[S. Chowdhury]{Department of Mathematics, The Ohio State University. Phone: (614) 292-6805.}
\address[B. Dai]{Department of Computer Science, Dartmouth University.}
\address[F. M\'emoli]{Department of Mathematics and Department of
  Computer Science and Engineering, The Ohio State University. Phone: (614) 292-4975,
Fax: (614) 292-1479.}

\begin{abstract}
We develop of a line of work initiated by Curto and Itskov towards understanding the amount of information contained in the spike trains of hippocampal place cells via topology considerations. Previously, it was established that simply knowing which groups of place cells fire together in an animal's hippocampus is sufficient to extract the global topology of the animal's physical environment. 
We model a system where collections of place cells group and ungroup according to short-term plasticity rules. 
In particular, we obtain the surprising result that in experiments with spurious firing, the accuracy of the extracted topological information decreases with the persistence (beyond a certain regime) of the cell groups. This suggests that synaptic transience, or forgetting, is a mechanism by which the brain counteracts the effects of spurious place cell activity. 
\end{abstract}
\maketitle
\date{\today}

\tableofcontents

\section{Introduction}

The premise of our work is the commonly accepted theory that an animal's awareness of its surroundings---the physical \emph{stimulus space}---is encoded in the firing activity of \emph{place cells} that are predominantly found in the CA1 and CA3 regions of its hippocampus \cite{neuro-book}. Place cells are characterized by having firing patterns that are restricted to spatially localized regions called \emph{place fields} \cite{o1971hippocampus}. 
Experimental results \cite{brown1998statistical} suggest that the firing patterns or \emph{spike trains} of place cells contribute spatial information that the brain uses to infer properties of the stimulus space. 
This has led to some interest in the following question: Without assuming place field information, what information can be extracted from only the spike trains of place cells? In this paper, we restrict the context of this question to rodent place cells. In \cite{curto2008cell}, the authors used a mathematical shape analysis tool called \emph{homology} to count the number of obstacles in an arena being explored by a rodent. 
This idea was further developed in \cite{dabaghian2009topological, dabaghian2012topological}, where the authors used a time-series extension of homology called \emph{persistent homology} (PH) \cite{frosini1992measuring, edelsbrunner2008persistent, ghrist2008barcodes, carlsson2009topology, edelsbrunner2014persistent} to identify bounds on the choices of parameters (e.g. number of place cells, sizes of place fields, firing rate) with respect to which homology could correctly identify topological features (i.e. connectivity/adjacency of locations) of the stimulus space from spike train data. Specifically, the authors simulated spiking activity of rodent place cells as the animal explored arenas having one or two obstacles (the topological features). The ground truth to be recovered was the number of obstacles in each arena. In \cite{cosyne}, this study was expanded to utilize a newly developed method called \emph{directed network persistent homology} \cite{dowker-arxiv}, which significantly improved the 1-nearest neighbor classification error from that obtained via the methods used in \cite{dabaghian2012topological}. 

\begin{wrapfigure}{r}{0.5\textwidth}
  \begin{center}
   \captionsetup{width=.9\linewidth}
    \includegraphics[width=0.48\textwidth]{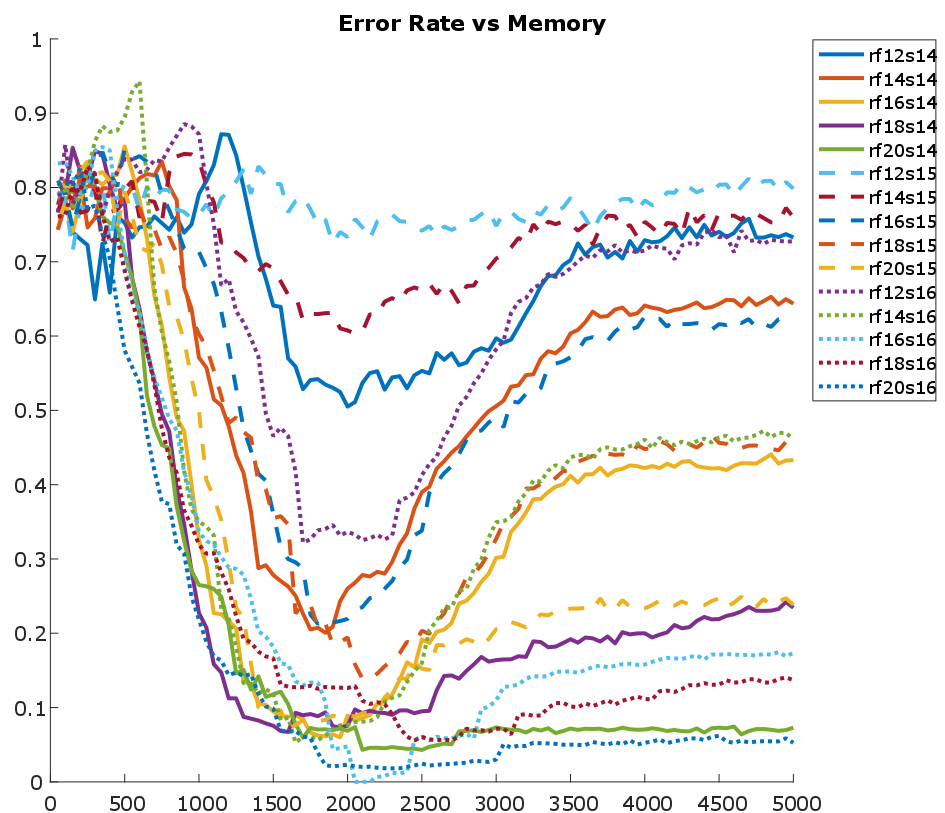}
    \medskip
  \caption{Our results at a glance: a plot of error rate against memory size $\t$. The maximum $\t$ is 5000. This represents the full duration of the experiment. \textbf{Legend}: numbers following ``rf"  and ``s" are mean firing rates (in $\hz$) and place field sizes (in $\cm$), respectively.}
  \label{fig:error-rates-all}
    \end{center}
    
\end{wrapfigure}

The current thrust of our work is to incorporate the notion of \emph{forgetting} information in the model for memory. 
The main tool that we are using for our analysis is \emph{zigzag persistent homology} \cite{zigzag}, which allows the user to discard information (i.e. \emph{forget}) in a systematic way while still building a persistent homology signature. Our current model for forgetting is very simple: for a specified length of time $\t$ that is less than or equal to the experiment duration, we specify that the rodent remembers, at time $t$, only the information that it received in the window $[t-\t,t]$. We test this forgetting model for a range of 100 different values of $\t$, on a dataset consisting of 750 simulations produced from a variety of firing rate and place field size parameters.

Our results support the surprising concept that \emph{more memory is not always better}. A plot of the 1-nearest neighbor classification error shows that as $\t$ increases from 0, the error rate initially drops, then achieves a minimum before rising again. Thus it seems that to achieve the best possible result in ``learning" an arena, the rodent needs a balance between remembering and forgetting information. This result, while nonintuitive, is in line with recent literature in neuroscience \cite{richards2017persistence} where it has been proposed that forgetting is an important step in the learning process. 

\paragraph*{Our premise.}

We restrict our study to the place cell activity in a rodent as it explores a planar region containing some obstacles. We refer to this region as an \emph{environment} or an \emph{arena}.

When referring to the \emph{topology of the environment}, we specifically refer to a mathematical shape descriptor called \emph{homology}. Homology has different interpretations in each dimension; 0-dimensional homology of a space refers to the number of its connected components, 1-dimensional homology refers to the number of loops, and so on (see \S\ref{sec:ph}). In this paper, we focus only on 1-dimensional homology. The environments we consider are square regions with obstacles that the rodent cannot pass through; the 1-dimensional homology of each such environment is the number of obstacles contained in it, and this is the number of interest.

Next we describe what we mean by the \emph{topology of the synaptic potentiation complex}. By potentiation, we simply mean increased (above baseline) synaptic connectivity. At the biological level, one of the contributors to this effect in the CA1 region of the hippocampus has been characterized as changes to AMPA receptors in the postsynaptic membrane (e.g. through an increase in channel conductance or in the number of receptors) \cite{benke1998modulation, plant2006transient}. These changes are continuous, but for the purposes of the mathematical model, we will simplify the neurobiological effects to discrete ``increase/stay at baseline" events. 

Suppose we are conducting an experiment where we track the activity of an ensemble $S$ of place cells as an animal explores one of the regions described above. For any subset $\s$ of $S$ containing at least two cells, any time $t$ (a time instance during the experiment) and a time interval $\t$, we define the following potentiation function:
\[A_\t(\s,t) \defeq \begin{cases}
1 & : \text{ if the cells in $\s$ cofired in the time interval $[t- \t,t]$}\\
0 &: \text{ otherwise}
\end{cases}\]
Here we write \emph{cofiring} to mean that \emph{all} the cells in $\s$ fired above a threshold (see \S\ref{sec:methods} for our threshold choice) in a window of two theta cycles ($\sim$350ms). In \cite{curto2008cell}, $\s$ was referred to as a \emph{cell group}. This potentiation function is motivated by the classic Hebbian ``fire together, wire together" principle. However, if enough time passes during which cells that have ``wired together" have not fired together again, then they ``unwire".

Using this potentiation function with supplied values of $t$ and $\t$, one builds a \emph{dynamic simplicial complex} (see \S\ref{sec:ph}) with node set $S$ and a simplex for each subset $\s$ with activation 1. We call this dynamic simplicial complex the \emph{synaptic potentiation complex}. 
More specifically, suppose that the experiment starts at time $0$ and ends at time $T$, and that data points are recorded at times $\{0,1,2,\ldots, T-1,T\}$. Fix a value of $\t$. Then a simplicial complex can be built via the rule given above for each value of $t \in \{0,1,2, \ldots, T\}$. Thus we obtain a sequence $\{K_i\}_{i=0}^{5000}$ of simplicial complexes. The point to note here is that this sequence is a bona fide dynamic simplicial complex: simplices that are ``old" (representing place cells that have not cofired recently) are dropped from the complex, even as new simplices are being added. The lifetime of a simplex is controlled by the $\t$ parameter, which we interpret as the \emph{memory capacity} of the animal. Our work can broadly be described as computing the topology of this dynamic complex for differing values of $\t$ to study how memory affects the animal's understanding of the topology of the environment.

The main mathematical intuition underlying our work is a result from algebraic topology called the \emph{Nerve Theorem} \cite{bjorner1995topological}, which states that if a space $X$ can be decomposed into smaller subspaces $\{A_1,\ldots, A_n\}$ satisfying some well-behaved intersection properties, then the topology of the space is equivalent to the topology of a simpler space called the \emph{nerve} of the decomposition. The nerve is a simplicial complex built on the indexing set $I \defeq \{1,\ldots, n\}$ as follows: a subset $\s \subseteq I$ belongs to the nerve if and only if the intersection $\cap_{i \in \s}A_i$ is nonempty. 
The crucial observation made by Curto and Itskov \cite{curto2008cell} is that the topology of the synaptic potentiation complex can be used to extract the topology of the environment, by virtue of the Nerve Theorem. This can be seen as follows (for simplicity, assume $\t = T$): suppose that we have a labeled collection of $n$ place cells whose place fields $A_1,A_2,\ldots, A_n$ cover all of the accessible regions of an environment (so obstacles are not fully covered). 
Suppose also that all the place fields are convex. The Nerve Theorem then guarantees that the nerve simplicial complex associated to the place fields has the same topology as the environment. Assuming that cell groups that cofire correspond precisely to place field intersections, it then follows that the topology of the environment is the same as the topology of the potentiation complex. 

\subsection{Related literature and our contributions}

The ideas behind this project were conceived in the context of the 2014 NSF IIS-1422400 grant. An offshoot of the same grant was released earlier this month as \cite{babichev2017robust}. Both of these papers use zigzag persistent homology, but the conclusions drawn in the two works are different. In fact, they address separate research objectives that were both contained in the proposal for the IIS-1422400 project.

\subsubsection{Tracking topological changes via persistent homology}
\label{sec:past-work-ph}
In \cite{dabaghian2012topological}, a similar topological framework was used to understand the following question: how long does it take for the topology of an animal's synaptic potentiation complex to achieve the topology of its environment? At the beginning of the experiment, it is assumed that the animal has just begun exploring the environment, and that new connections are being added to the potentiation complex. As the experiment runs and the animal explores the environment, the topology of the network changes. The authors of \cite{dabaghian2012topological} used a relatively new computational technique---\emph{persistent homology}---to track these changes and identify the first time that the topology of the potentiation complex matches the topology of the environment in a stable manner. This event was referred to as \emph{learning}, and the subspace of parameter space (number of place fields, place field size, mean firing rate) in which learning occurred was identified and named the \emph{stable learning region}. In \cite{dabaghian2012topological}, it was argued that persistent homology provides more information than its non-persistent counterpart (that we call \emph{flat homology}). 

We made a further advance on this idea of using persistent homology in \cite{cosyne}, where we first encoded the place cell firing information into a directed, weighted network. The nodes in this network were the place cells, and the asymmetric weight between a pair $(i,j)$ of cells was given by the relative frequency with which cell $j$ fired in a brief time window after cell $i$ had already fired. This particular encoding thus captured causality relations between the firing activity of the place cells. After producing these networks, we used a notion of \emph{directed network persistent homology} developed in \cite{dowker-arxiv} to obtain persistence barcodes. We then compared these barcodes to each other via a natural metric called the \emph{bottleneck distance}. Using this distance, we computed the 1-nearest neighbor classification error rate and analyzed the hierarchical cluster structure via single linkage dendrograms. The objective of computing these error rates and dendrograms was to show that the network persistent homology method could better distinguish between the barcodes arising from different types of arenas than the PH method used in \cite{dabaghian2012topological}. As we describe in \S\ref{sec:contributions} and \S\ref{sec:methods}, our analysis methods in this paper follow those used in \cite{cosyne}.

For complementary work on neural topological representations (using persistent homology and other techniques), see \cite{chen2014neural,giusti2015clique, spreemann2015using, curto2017can}. 

\begin{figure}
\includegraphics[width = \textwidth]{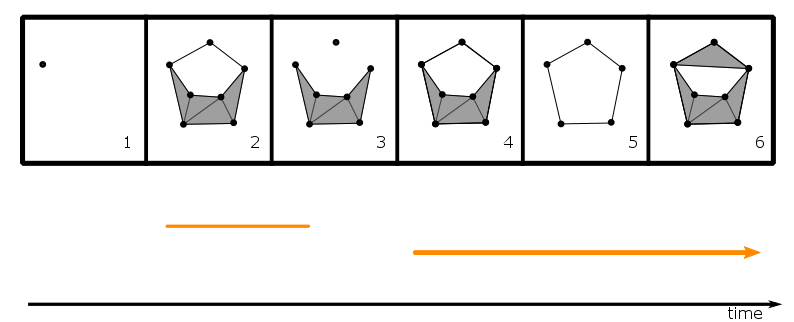}
\caption{An example of a dynamic simplicial complex and its 1-dimensional barcode (see \S\ref{sec:zigzag}). The simplicial complexes in frames 1 and 3 include into 2; 3 and 5 include into 4, and 5 includes into 6. Any dynamic simplicial complex is characterized by this property that for any two simplicial complexes occurring consecutively in the sequence, one is contained in the other (in some order).}
\label{fig:zz-toy}
\end{figure}

\subsubsection{Topological changes in the presence of decaying connections}

In \cite{babichev2017robust} the authors considered rodent hippocampal firing models that are related to ours, but different in key aspects. Here the authors considered one square arena with one square hole in the middle, and one trajectory that fully explores the arena. In their model, the authors considered only pairwise connections between neurons. These pairwise links were added by a rule similar to the cofiring rule described above. The links were allowed to decay via two models: an exponential decay model, and a fixed lifetime model similar to ours. A simplicial complex was then obtained from these pairwise links by taking the associated \emph{clique complex}. In this construction, a $d$-simplex for $d\geq 2$ is added whenever all the edges between the vertices in the simplex are present. This is less expensive than constructing a nerve complex. However, the clique complex is generally different from the nerve complex, and so the guiding principle afforded by the Nerve Theorem does not apply in this setting. 

The authors of \cite{babichev2017robust} applied zigzag persistent homology to this dynamic clique complex for a range of decay rates and obtained persistence barcodes. The crux of their work was in showing that at least for the exponential decay model, the 0 and 1-dimensional homology vector spaces both have rank 1 for all times after some time threshold. In other words, they showed that the instantaneous 0 and 1-dimensional homology rank matches the 0 and 1-dimensional homology rank of the arena, after some time length.

\subsubsection{Our approach: zigzag persistent homology, bottleneck distances, and error rates}
\label{sec:contributions}

The standard persistent homology setup as described above is unable to accept as input the fully dynamic simplicial complexes
constructed via $A_\t$. The recourse is to consider the more powerful tool of zigzag persistent homology \cite{zigzag}, which performs persistent homology computations on dynamic simplicial complexes (see Figure \ref{fig:zz-toy} for an illustration). 
Our chief goal is to enrich the existing literature by testing the intuitive hypothesis that the more ``memory" the place cells have (i.e. the greater the value of $\t$), the more accurately the topology of the synaptic potentiation complex matches the topology of the environment.

\begin{wrapfigure}{R}{0.4\textwidth}
  \begin{center}
    \includegraphics[width=0.4\textwidth]{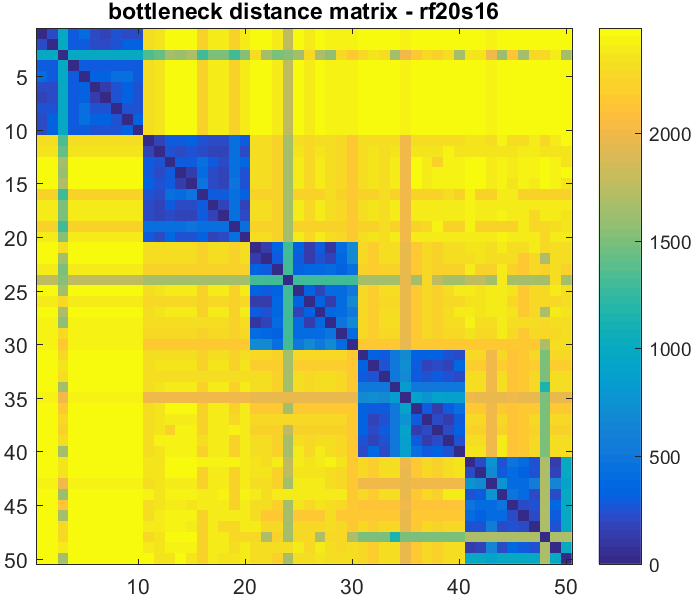}
  \caption{A bottleneck distance matrix obtained with mean firing rate $20\hz$, place field size $16\cm$, and $\t=2000$. Since the experiment duration is 5000, the maximum bar length in a barcode is $\leq 5000$. The bottleneck distance between barcodes with a different number of long bars is approximately $5000\times1/2 = 2500$ (shown in yellow).}
  \label{fig:bottleneck}
    \end{center}
    
\end{wrapfigure}

\begin{figure}
\includegraphics[width = 1\textwidth]{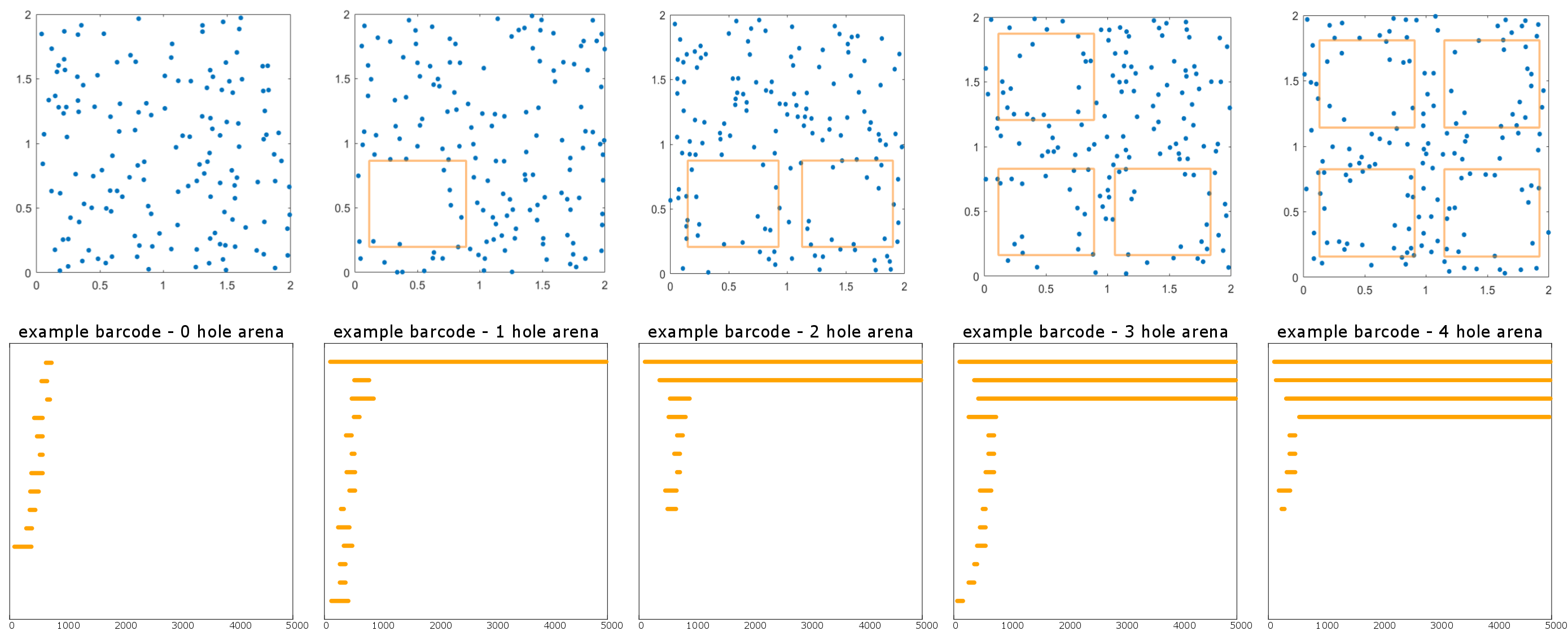}
\caption{\textbf{Top row}: the five types of arenas we considered. Each arena has dimensions $200\cm\times 200\cm$. Blue dots represent centers of place fields. Orange squares mark positions of circular obstacles of radius $25\cm$. \textbf{Bottom row}: Persistence barcodes obtained from running simulations on the five different arenas with mean firing rate $20\hz$, place field size $16\cm$, and $\t=2000$. Notice that the number of long bars in each barcode is equal to the number of obstacles in the corresponding arena.}
\label{fig:barcode}
\end{figure}

To this end, we chose a set of biologically plausible \cite{dabaghian2012topological} parameters for mean firing rates ($\{12\hz$, $14\hz$, $16\hz$, $18\hz$, $20\hz\}$) and place field radii ($\{14\cm,15\cm,16\cm\}$).
For each of the 15 choices of firing rate-place field size pairs, we generated 10 simulations of a rodent's trajectory around an arena with $0,1,2,3$ and $4$ obstacles for a total of $15\times 10\times 5=750$ datasets. For each dataset, we computed the synaptic potentiation function $A_\t$ for 100 choices of $\t$. These choices were obtained as follows: the experiment duration (10 minutes) was divided into $5000$ time bins, and the values of $\t$ were in the range $[50,5000]$ in increments of $50$. Thus we obtained $750 \times 100 = 75000$ dynamic simplicial complexes. 
Next we applied zigzag persistent homology to these filtrations to obtain 75000 \emph{persistence barcodes} (see \S\ref{sec:ph}). Each of these persistence barcodes provides a visual representation of the time intervals during which the topology of the potentiation complex matches the topology of the environment. Some example barcodes are provided in Figure \ref{fig:barcode}.

Our next step was where we differed completely from approaches taken in \cite{curto2008cell, dabaghian2012topological, babichev2017robust}, but remained similar in spirit to our prior work in \cite{cosyne} (see \S\ref{sec:past-work-ph}). The output space of persistent homology (often called the \emph{barcode space}) is equipped with a natural metric called the \emph{bottleneck distance}. 
For each parameter pair and each choice of $\t$, the pairwise bottleneck distance matrix between the corresponding 50 persistence barcodes determines how they are clustered together (see Figure \ref{fig:bottleneck}). Intuitively, there should be five clusters, corresponding to each type of arena. We computed a $50\times 50$ pairwise bottleneck distance matrix for each choice of $\t$ and parameter-pair, and performed 1-nearest neighbor classification on each of these datasets. The corresponding error rate provides a measure of the separation of these clusters in barcode space. 
We plotted the error rate as a function of $\t$ for each of the 15 parameter choices (see Figure \ref{fig:error-rates-all}). 
Our computational results (see Figure \ref{fig:error-rates-all} and \S\ref{sec:results}) contradict the seemingly natural ``more memory is better" hypothesis. Moreover, our results also indicate that the error rate achieves a global minimum at approximately $\t = 2000$, and then increases as $\t$ increases.
This suggests that forgetting connections---i.e. \emph{transience}---plays an important role in learning the topology of the environment.

In the pipeline described above, it was not necessary to directly examine each of the computed barcodes. However, as we describe next, there is valuable information that can be obtained by direct inspection of individual barcodes. The key principle motivating the following discussion is that the long bars in a barcode correspond to meaningful topological features, whereas short bars correspond to topological noise \cite{carlsson2009topology}. Consider the persistence barcodes in Figure \ref{fig:barcode}. These are example barcodes obtained from one of our simulations using mean firing rate $20\hz$, place field size $16\cm$, and $\t=2000$. Assume for now that a bar is ``long" if its length is $\geq 4000$. Then the number of long bars in each barcode is equal to the number of obstacles in the corresponding arena. This difference between the barcodes will be sharply reflected in the bottleneck distance, as can be seen in Figure \ref{fig:bottleneck}. Thus there are meaningful topological characteristics at play behind the classification via bottleneck distances.

While analyzing the increase in error rate shown in Figure \ref{fig:error-rates-all}, we found that it was useful to plot the number of long bars in a barcode as a function of $\t$. The complication in this strategy is that for large numbers of simulations (which arise from a set of stochastic processes), it is not straightforward to pick a decision boundary for counting a bar as ``long" or otherwise. We settled on $\ell\defeq 4000$ as the decision boundary because for the barcodes we examined, it seemed that each obstacle was giving rise to an individual bar of length $\geq 4000$.

\subsubsection{Transience in the hippocampus}

Each of the models of \cite{curto2008cell} and \cite{dabaghian2012topological} considered synaptic potentiation events and ignored depotentiation. So for each trial simulation in each of these models, all of the cofiring events that occurred in the duration of the trial were used in informing the topological inference. 

In our interpretation of place cell cofiring as synaptic potentiation, using full knowledge of cofiring has the following intuitive meaning: the animal is assumed to have full \emph{memory} of the cofiring events in its hippocampus. Thus increased synaptic potentiation is interpreted as \emph{memory}. Biologically, this interpretation is well-documented \cite{josselyn2015finding, tonegawa2015memory}. It is important to note that depressed synaptic connections also play a role in memory \cite{dudek1992homosynaptic, kemp2007hippocampal}, perhaps by maximizing signal-to-noise ratio \cite{dayan1991optimising}, but for simplicity, we will associate memory with only increased synaptic potentiation, i.e. by a value of 1 for the $A_\t(\s,t)$ function above. Given the interpretation of memory as increased synaptic potentiation, it follows that forgetting, or \emph{transience}, involves destabilization of synaptic connections (e.g. via depotentiation or elimination of synaptic connections) \cite{richards2017persistence}.  

The biological mechanisms responsible for natural forgetting have been documented in \cite{richards2017persistence}; we summarize some of those here. For example, just as an increase in the number of AMPA receptors in the postsynaptic membrane is associated with increased memory, a decrease in this number via AMPA receptor endocytosis has recently been associated with forgetting \cite{dong2015long}. 
Other work has shown that natural forgetting involves a protein called Rac \cite{shuai2010forgetting, liu2016hippocampal}---overexpression of Rac accelerates 
forgetting, and inhibition extends memory. On a longer timescale (thus less relevant to our work, but still interesting), another contributor to transience is hippocampal neurogenesis: the generation of neurons from stem cells in the dentate gyrus region of the hippocampal formation \cite{nicolas2007synapse, toni2008neurons}. It has recently been shown that this generation process is competitive, and that the adult-born neurons may remap preexisting connections in the hippocampus \cite{mcavoy2016modulating}. 

\subsubsection{Transience and overfitting}

It has been proposed \cite{richards2017persistence} that the biological motivation for transience is at least twofold:
\begin{itemize}
\item[(1)] forgetting old memories allows the brain to remove outdated information, thus improving behavioral flexibility, and 
\item[(2)] forgetting particular details and remembering only sparse representations of ideas or concepts helps the brain avoid ``overfitting" when performing prediction tasks on new data points. 
\end{itemize}
Overfitting is a term also used in the statistical learning literature to describe a model that has far too many parameters in relation to the amount of available data \cite{friedman-book}. The problem with any such model is that while it may fit a particular dataset very well, it will often be very poor at fitting a new dataset, even if the new data follows the same distribution as the training data. In the artificial neural networks literature, numerous techniques have been developed to reduce overfitting. One technique, named ``optimal brain damage", involves training the network and then systematically setting some weights to zero \cite{lecun1989optimal}. Another technique, called dropout, involves randomly dropping neurons and their connections from the network during training \cite{srivastava2014dropout}. In \cite{richards2017persistence}, parallels were drawn between these techniques used in artificial neural networks and transience events that occur in the brain, and it was suggested that transience is a natural process that helps the brain avoid overfitting. 

Returning to our topological paradigm, we are faced with the following question: what does it mean to overfit a topology? A formal answer to this question is interesting in its own right, but we were more interested in the parallels between some of the techniques described above (such as zeroing out weights in an artificial neural network) and our definition of the synaptic potentiation complex (removing certain simplices instead of retaining all the activated simplices). In \cite{richards2017persistence}, it was suggested that transience is a mechanism which aids the brain in decision-making---the analogy with our computational findings is that transience aids in minimizing the error rate. 
\section{Results}
\label{sec:results}

\begin{figure}
\includegraphics[width=\textwidth]{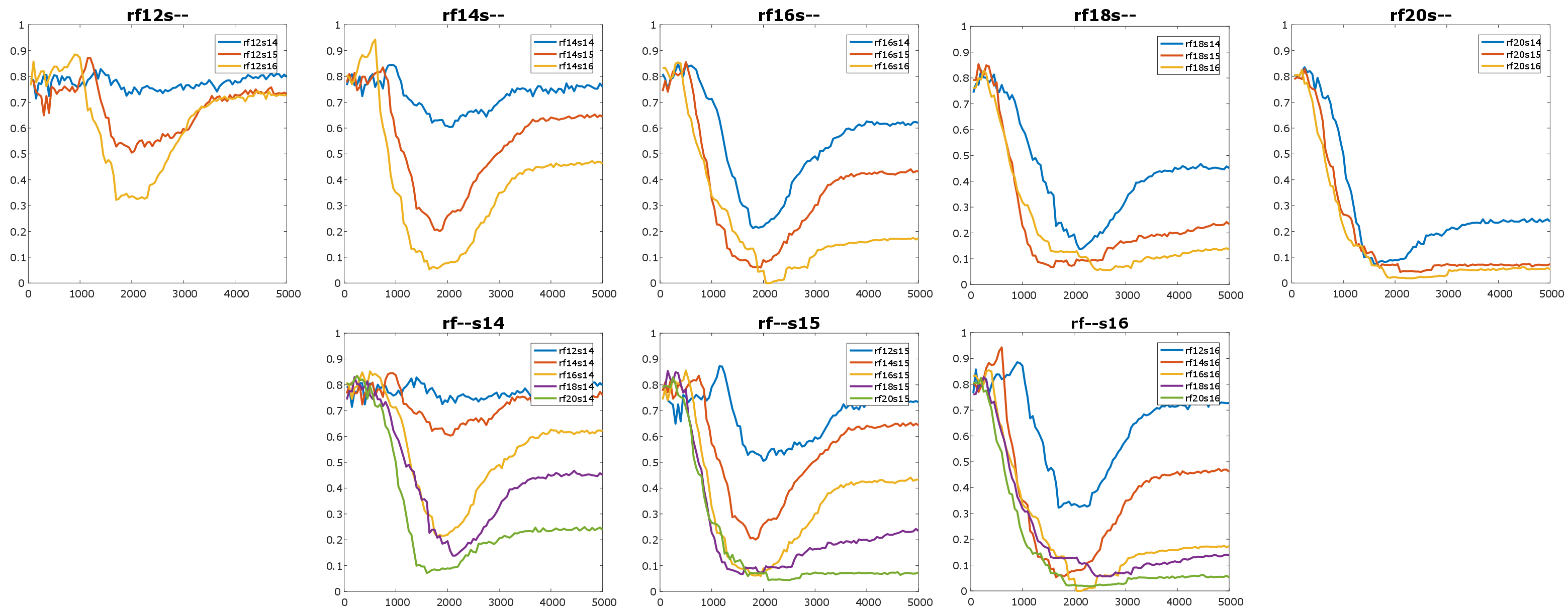}
\caption{Plots of error rates vs $\t$. \textbf{Top row:} Plots where the mean firing rate is fixed at $12\hz$, $14\hz$, $16\hz$, $18\hz$, and $20\hz$, respectively. Blue, red, and orange lines correspond to place field sizes of $14\cm$, $15\cm$, and $16\cm$, respectively.  
\textbf{Bottom row:} Plots where the place field sizes are fixed at $14\cm$, $15\cm$, and $16\cm$, respectively. Blue, red, orange, violet, and green lines denote mean firing rates $12\hz$, $14\hz$, $16\hz$, $18\hz$, and $20\hz$, respectively. As described in \S\ref{sec:results-error-param}, for each fixed value of $\t$, the overall error rate tends to decrease as the mean firing rate and place field size increase. 
}
\label{fig:error-rates}
\end{figure}

In Figure \ref{fig:error-rates}, we show the main results that we obtained. In five of the eight plots, we fixed the firing rate and plotted the error rate vs $\t$ for the three choices of place field size. In the remaining plots, we fixed the place field size and plotted the error rate vs $\t$ for the five choices of firing rates. In each of these figures, we see that the error rate drops to a minimum at $\t \approx 2000$.

For each error rate computation, we have a $50\times 50$ distance matrix of pairwise bottleneck distances between persistence barcodes obtained for each choice of $\t$ and parameter pair. If there is no real clustering in the 50 points in barcode space, then the classification rule is expected to be correct 1/5 times, thus giving an error rate close to $4/5 = 0.8$. 
If the dataset is tightly clustered into five well-separated regions, then the error rate is expected to drop to 0. 

\subsection{Stochastic firing increases error rate}

Why does the error rate decrease and then increase again, instead of simply dropping as memory ($\t$) increases? There are numerous complications surrounding this question: a priori, the trajectory model, the place field distribution inside each arena, the choice of firing model, and the choice of parameters can all play a role in shaping the error rate plot. 

In our initial experiments, we obtained this error rate phenomenon using a Poisson firing model and a random walk model for the rodent's trajectory. As our first step in understanding the results, we repeated all our simulations with trajectories given by a modified \emph{billiards model}, where the rodent followed a piecewise-linear trajectory and bounced off of walls. 
Some modifications were applied to ensure that the rodent visited corners obscured by obstacles with sufficient frequency (details in \S\ref{sec:methods}). The billiards model was chosen to ensure that the rodent would fully explore the environment with a reliable frequency. This prevented simulations where the rodent would explore the environment fully in a selected interval of time steps, and never again in the remaining time steps. The results in this paper are all obtained using the billiards model. However, switching from the random walk model to the billiards model did not 
alter
the error rate phenomenon---again we observed that the error rate would decrease, and then increase again. 

\begin{figure}
\includegraphics[scale=0.2]{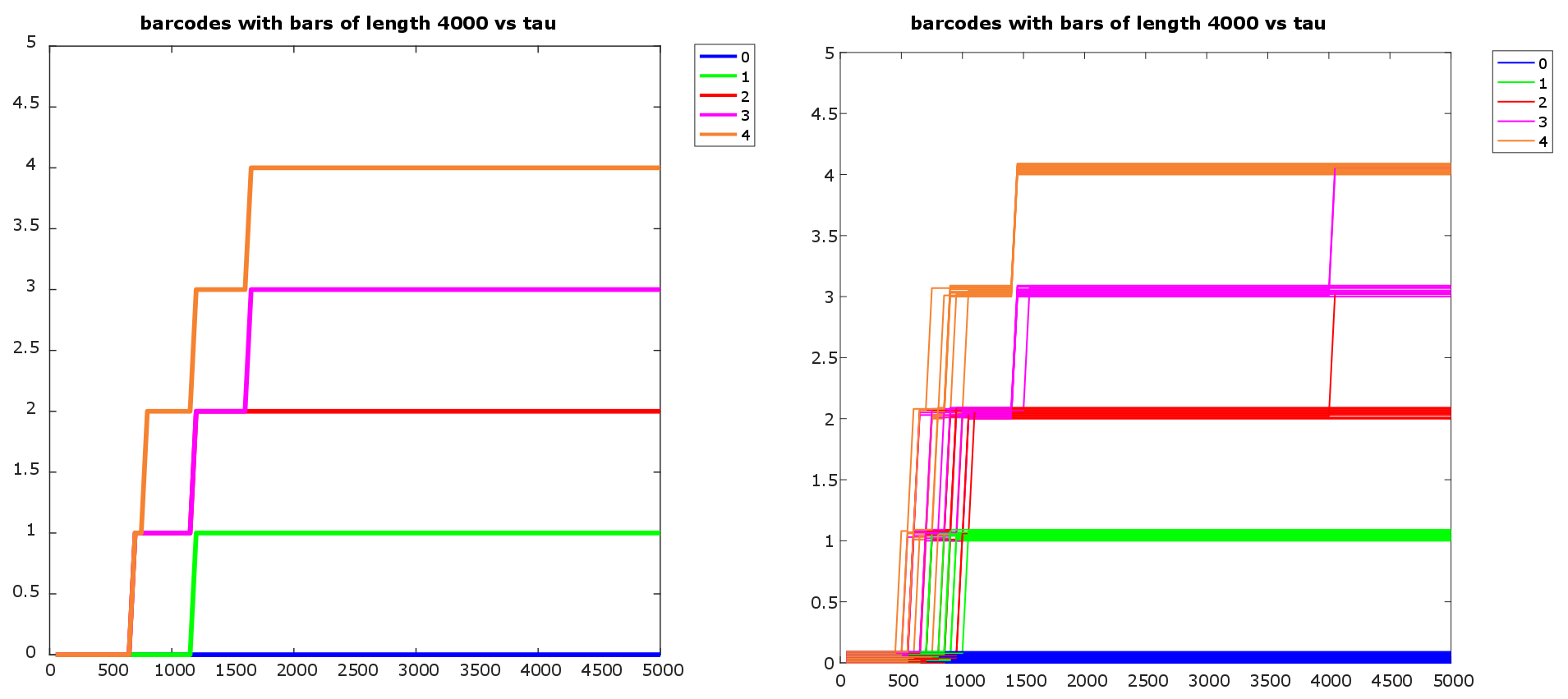}
\caption{Number of bars of length greater than $\ell = 4000$ in the barcodes obtained via the binary model (\textbf{left}) and the fuzzy binary model (\textbf{right}) as a function of $\t$. The plot on the left panel has only five lines. The plot on the right panel has 50 lines, 10 for each color (corresponding to 10 simulations on each arena). }
\label{fig:bar-4000}
\end{figure}

To fully simplify our model and protect against any procedural errors, we next replaced the Poisson firing model by a binary firing model where a place cell fired if and only if the rodent entered the corresponding place field (see \S\ref{sec:methods}). 
We ran this simulation using mean firing rate $20\hz$ and place field size $15\cm$.
Since there was no stochasticity in this model, we obtained 5 persistence barcodes for each choice of $\t$ (as opposed to $50$). In this model, there was no classification to be done and thus no error rate to calculate. However, we were still able to analyze the barcodes and verify that the zigzag persistence machinery was working correctly. 
In Figure \ref{fig:bar-4000} (left panel) we plotted the number of bars of length greater than $\ell  \defeq 4000$ in each of the five barcodes as a function of $\t$. We chose $\ell=4000$ because in our simulations using the billiards model, the rodent fully explored each environment within the first 1000 time steps. 
In this plot, we see that for $\t \approx 1600$ and above, there is exactly one barcode having $i$ bars of length $\geq \ell$, for each $i = 0,\ldots, 4$. This is expected behavior---as discussed in \S\ref{sec:contributions}, the number of long bars in a barcode is a count of the number of obstacles in the environment that produced the barcode. In particular, the barcodes \emph{do not} overcount or undercount the correct number of obstacles in the environment for any $\t$ value greater than $\approx 1600$.

Next we introduced some stochasticity into the binary firing model and considered the \emph{fuzzy binary firing model} (details in \S\ref{sec:methods}). In this model, a place cell fired deterministically if the rodent entered its place field, and stochastically if the rodent was within some bounded distance from the place field center. This fuzzy model is a compromise between the fully deterministic binary firing and fully stochastic Poisson firing models. We ran 10 simulations of this fuzzy model with mean firing rate $20\hz$ and place field radius $15\cm$. In Figure \ref{fig:bar-4000} (right panel), we plotted the number of bars of length greater than $\ell = 4000$ in the 50 barcodes as a function of $\t$. The important observation about this figure is that for large values of $\t$, some of the barcodes are \emph{overcounting} the number of obstacles in the corresponding environment. Thus we expect the error rate to increase for large $\t$ values. This suggests that stochastic firing causes error rate to increase for large values of $\t$.

\subsection{An explanation for the error rate dip}

\label{sec:results-error-dip}

\begin{figure}
\includegraphics[width = 0.9\textwidth]{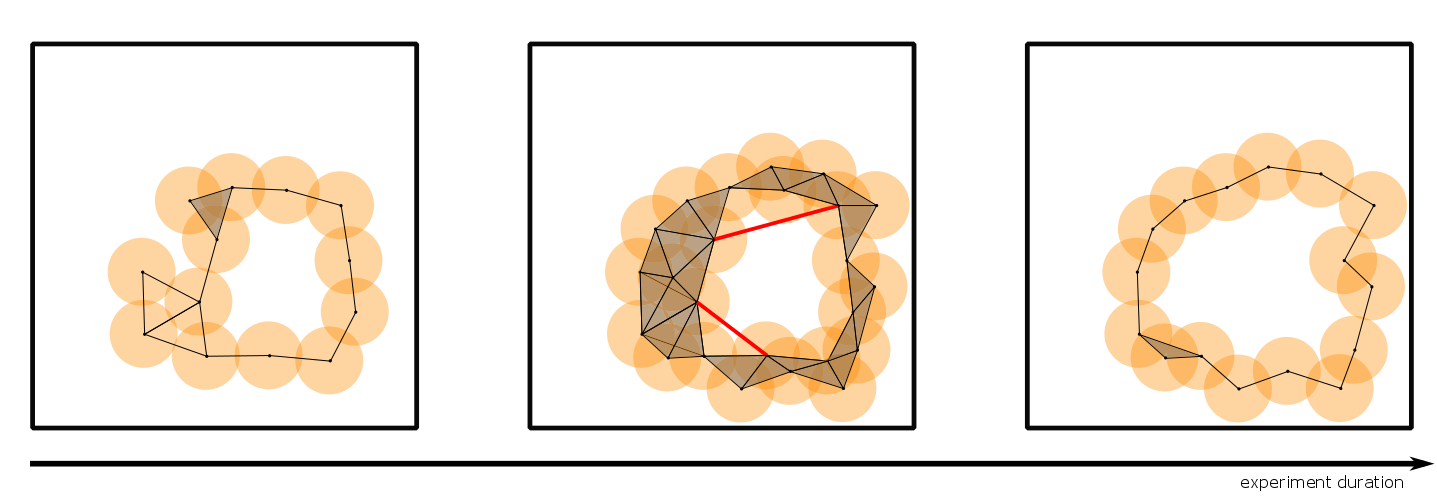}
\caption{The dynamic simplicial complex described in \S\ref{sec:results-error-dip}. The two triangular loops in the left panel are examples of \emph{off-obstacle loops} (see \S\ref{sec:results-error-dip}). Red lines in the middle panel represent spurious connections that are formed by stochastic activity. The loops containing these red lines are examples of \emph{on-obstacle loops} (\S\ref{sec:results-error-dip}). 
}
\label{fig:error-explanation}
\end{figure}

The most interesting aspect of our results is the dip in the error rate curves of Figure \ref{fig:error-rates-all}. There is an intuitive explanation for this dip, which we illustrate in Figure \ref{fig:error-explanation}. Here we depict a 1-obstacle arena.
The orange disks represent place fields centered around the obstacle, and the overlaid dynamic simplicial complex is the synaptic potentiation complex. Since a place cell fires with high probability when the rodent is in its place field, a pair or triple of place cells is likely to cofire when their place fields intersect. Thus we have drawn edges (in black) between the centers of pairs of intersecting place fields, and added faces (in grey) between triples of intersecting place fields. 

In addition to the events described above, the Poisson firing model (or the fuzzy binary firing model) both permit place cells to cofire even when their place fields do not intersect. Two of these instances have been marked by red lines in the middle figure. Now we turn to the question of counting loops in each of these simplicial complexes. There is one principal loop corresponding to an obstacle in each of the figures. In the left and middle figures, there are two additional loops in black and red, respectively. 

The short black loops do not enclose any part of an obstacle, so we call them \emph{off-obstacle loops}. As the rodent explores the environment, it will pass through these loops and activate more place fields that ``fill in" these short loops. This is illustrated in the middle figure. The red loops, which we call \emph{on-obstacle loops}, cannot be filled in this way, simply because the animal cannot pass through the obstacle. There are two plausible methods by which the red loops can be filled in:
\begin{itemize}
\item[(1)] enough triples of place cells may cofire stochastically to fill in the loop, and 
\item[(2)] the red edge forming the loop may simply be forgotten after some time. 
\end{itemize}
Given that cofiring between cells having non-intersecting place fields is a rare event, the second method seems to be more appropriate for explaining how and when the red loops disappear. 

Accepting the ``forgetting a loop" explanation, it is suggestive to consider the relationship between $\t$ and the on-obstacle loops. The greater the value of $\t$, the longer an on-obstacle loop remains in the simplicial complex. The longer an on-obstacle loop remains in the complex, the more it contributes (erroneously) to the persistence barcode. Thus having a large value of $\t$ can indirectly increase the error rate. 

At the other extreme, suppose that the value of $\t$ is taken to be very small. In this situation, there may never be enough memory for the dynamic simplicial complex in Figure \ref{fig:error-explanation} to contain the principal loop. Thus very small values of $\t$ filter out too much information for the dynamic complex to have any meaningful topological information.

\subsection{Effect of parameters on error rate}
\label{sec:results-error-param}

We observe directly from Figure \ref{fig:error-rates} that increasing the mean firing rate and the place field size causes the error rate to decrease. The physical intuition behind this decrease is as follows: increased firing rate and place field size both contribute to increased cofiring. Increased cofiring in turn causes off-obstacle loops (as described in \S\ref{sec:results-error-dip}) to be filled in more quickly. Thus the (erroneous) contribution of these off-obstacle loops to the barcode is reduced. The bottleneck distance is then better able to distinguish between barcodes arising from different arenas, and the error rate consequently decreases. 

The 1-nearest neighbor classification error can be interpreted as a measure of the separation of the different clusters \cite{friedman-book}.  
In Figure \ref{fig:dendro-tau2000}, we plotted the single linkage dendrograms \cite{friedman-book} 
depicting the clustering behavior in the persistence barcodes arising from the 15 different choices of mean firing rate-place field size pairs. Here $\t$ is fixed at its experimentally determined optimal value of $2000$. Notice that the clustering improves significantly as the mean firing rate and place field size increase. 

\afterpage{
\begin{landscape}

\begin{figure}
\includegraphics[width=1.4\textwidth, keepaspectratio]{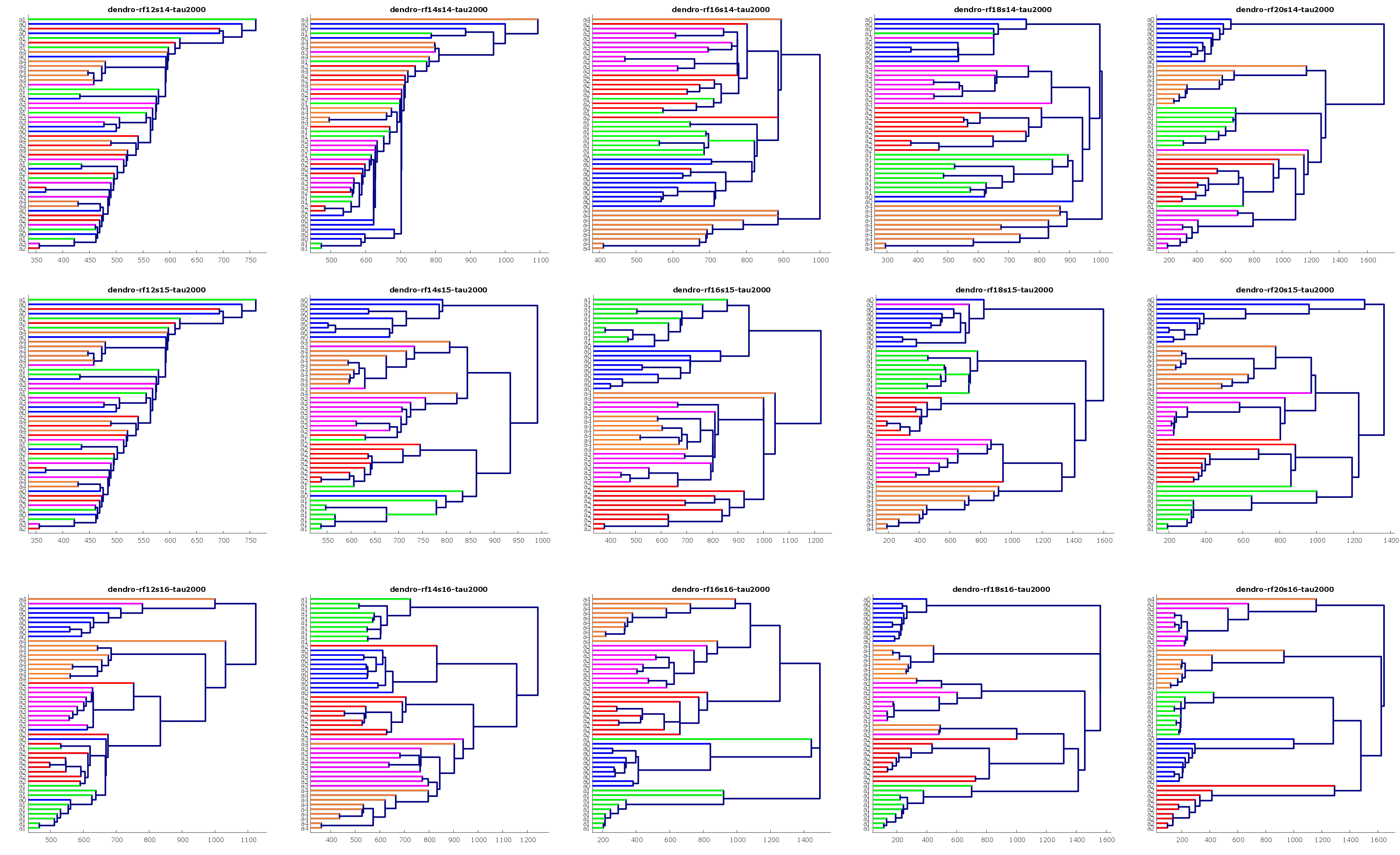}
\caption{Single linkage dendrograms showing clustering of persistence barcodes obtained from 15 pairs of parameter choices. 0-4 hole arenas correspond to blue, green, red, magenta, and orange lines, respectively. 
Top, middle, and bottom rows correspond to fixed place field sizes of $14\cm$, $15\cm$, and $16\cm$, respectively. From left to right, the columns correspond to fixed firing rates of $12\hz$, $14\hz$, $16\hz$, $18\hz$, and $20\hz$, respectively. See \S\ref{sec:results-error-param} and \S\ref{sec:methods}. 
}
\label{fig:dendro-tau2000}
\end{figure}

\end{landscape}
}

\section{Discussion}

Our results suggest that transience, or forgetting, is an important mechanism for learning the topological characteristics of an environment. This is in line with recent perspectives appearing in the neuroscience literature \cite{richards2017persistence} on the importance of forgetting for better decision-making. 

From our viewpoint, a useful next step would be to repeat our simulations using different decay models, e.g. the exponential decay model used in \cite{babichev2017robust}. The ultimate goal would be to procure experimental data and test the hypothesis that we put forth in this paper regarding transience and topological learning. There has already been some interest in producing experiments to verify the predictions arising from algebraic topology, e.g. \cite{giusti2015clique}. Of course, there are numerous difficulties belying the use of real data that we ignored for the purposes of simulation: a significant number of place fields may be non-convex, and the data may be highly noisy. Despite these obstructions, there appears to be no obvious drawback to using zigzag persistent homology and bottleneck distances on real data. The complexity of the calculations scales with the number of place cells, and the number of place cells that can be recorded for real data is well within the capabilities of the available zigzag software.

\section{Methods}
\label{sec:methods}

The following notation is used below:
\begin{itemize}
\item \numpf---Number of place fields
\item \sizpf---Size (radius) of each place field
\item \meanf---Mean firing rate
\item \firewin---Window over which cofiring occurs
\item \fireth---Firing threshold; cell is active if it fires above this quantity at any time
\end{itemize}

\subsection{Arena generation}

The family of arenas $\mc{A}:=\{A_0,A_1,A_2,A_3,A_4\}$ were generated as follows. Each $A_i$ consisted of a $200\cm \times 200\cm $ square with $i$ obstacles of radius $25\cm$. The obstacles were centered at coordinates $(50 \cm,50 \cm)$, $(150 \cm,50 \cm )$, $(50 \cm ,150 \cm )$, and $(150 \cm,150 \cm)$, respectively. The arena generation code was supplied with parameters \numpf{} and \sizpf{}. Each arena was generated with $\numpf$ place fields with centers scattered uniformly at random, subject to the following rules: no place field center was placed within $25\cm$ of any obstacle, and any two place field centers were at least $\sizpf{} \times 25\cm$ apart. The values used for \sizpf{} were $\{14\cm, 15\cm, 16 \cm\}$. The value of \numpf{} was fixed at $150$.

\subsection{Trajectory model}

The trajectory was generated via a billiards model that we now discuss. The starting position of the trajectory was chosen uniformly at random from the square $[50\cm,150\cm] \times [50\cm,150\cm]$, i.e. a square at the center of the arena. This point was initialized with a random initial direction. The trajectory then followed this direction until it intersected with a wall of the arena or an obstacle. If the trajectory intersected with a wall, it bounced off at the angle of reflection, with an error of 5 degrees (chosen from a uniform distribution). If the trajectory intersected with an obstacle, then two points were recorded: the point of entry $e$ and the projected point of exit $f$ that would occur if the trajectory could pass in a straight line through the obstacle. Next each point in the intersection of the trajectory and obstacle were mapped to the nearest boundary point of the obstacle. Thus the trajectory would meet the obstacle at $e$, then follow the boundary of the obstacle until reaching $f$, and then proceed at the initial angle  with which it had reached $e$.

The billiards model was chosen because it is known to be \emph{strongly mixing} \cite{sinai1970dynamical}, i.e. after some short period of time, a trajectory following the billiards model will have visited all regions of the arena uniformly. In our setup, it was important for the rodent to fully explore the arena, and to do so with some approximate periodicity during the course of the experiment.

The trajectory consisted of 5000 steps. The simulation was modeled after an experiment lasting 10 minutes. Each step was assumed to take $600s/5000 = 0.12s$. The speed of the rodent was fixed at $25\cm/s$.

\subsection{Firing models}

\paragraph*{Poisson}
For each arena-trajectory pair, a spike train was generated according to the firing model described next. At each time step $t$, a vector $\mathbf{f}^t$ was sampled from a lognormal distribution with mean \meanf{} and standard deviation $(1.2)\meanf{}$. Here $\mathbf{f}^t$ is a \numpf{}-dimensional vector of individual firing rate amplitudes, with one slot for each place field. We write $r_t$ to denote the position of the rodent at time $t$, and $\mathbf{r}$ to denote the list of place field centers. A vector of position-dependent firing rates was then calculated as follows:
\[\mathbf{r}^t:= \mathbf{f}^t \exp(-\norm{r - \mathbf{r}}/{2\sizpf{}^2}).\]
The firing model was assumed to be a Poisson process with mean given by $\mathbf{r}^t$ and time interval $\dt = 600/5000s = 0.12s$. At each time step, the number of spikes produced by each place cell was given by $\mathbf{s}^t:=\texttt{poissrnd}(\mathbf{f}^t\cdot\dt)$.

The following values of \meanf{} were used: $\{0.12\hz, 0.14 \hz, 0.16 \hz, 0.18 \hz, 0.20\hz\}$. The ensemble of spike trains is called a \emph{raster}.

While the Poisson firing model is meant to simulate real data, we found that the interpretation of results is easier by first considering some simpler firing models. We describe these models next. 

\medskip

\paragraph*{Binary}
In the \emph{binary firing model}, the raster was generated through a purely deterministic process. At each time step, a place cell generated a spike if the rodent's trajectory intersected the corresponding place field. The number of spikes at each time step was given by $\meanf\cdot \dt$. 

\medskip
\paragraph*{Fuzzy Binary} The \emph{fuzzy binary firing model} was a cross between the deterministic $\binary$ and stochastic $\poisson$ firing models. For each time step $t$, we considered the collection $A$ of place cells whose place fields intersected the rodent's trajectory at $t$, and the collection $B$ of place cells whose place field centers were within $2\sizpf$ of the rodent's trajectory. The cells in $A$ fired deterministically according to the binary model described above. The cells in $B$ activated with probability $0.2$, and once activated, they generated $\meanf \cdot \dt$ spikes.

We will write $\poisson$, $\binary$, and $\fuzzyb$ to denote the Poisson, binary, and fuzzy binary firing models, respectively.

\subsection{Memory model and the synaptic potentiation complex} 
The output of the preceding steps was an integer-valued matrix of dimensions $150 \times 5000$, with entry $(i,j)$ giving the number of times cell $i$ fired at time $j$. Next a list file was created which contained, for each time $t \in \{1,2, \ldots, 5000\}$, the index of each cell whose aggregate firing in a window of size \firewin{} starting at $t$ exceeded the threshold \fireth{}. 

From this list file, an event file containing \emph{addition} and \emph{deletion} events was generated. This is described next. At time 0, an empty simplicial complex $K_0$ was initialized. At time $t$, for $t > 0$, the list file was scanned to find all $k$-tuples of active cells, for $k \in \{0,1,2\}$. Such a $k$-tuple (i.e. a $k$-simplex) was added to $K_t$ if it was not already present in $K_{t-1}$. The deletion event was as follows: at each time $t$, the simplices that had appeared at time $t- \t -1$ and had \emph{not} appeared in the window $[t-\t,t]$ were deleted from $K_t$. 

The resulting object $\{K_t\}_{t=1}^{5000}$ was the synaptic potentiation complex. The special property of this dynamic simplicial complex was that it allowed for removal of simplices, i.e. the deletion events described above. The idea was that as the rodent explored and learned the arena, new simplices were added to the potentiation complex. Conversely, as time passed and the rodent forgot regions of the arena it had not visited recently, the corresponding simplices in the potentiation complex were removed. 

This dynamic simplicial complex was then passed into \texttt{Dionysus} \cite{morozov2012dionysus} for zigzag persistence computation, i.e. to compute the persistent homology of this at-times-growing, at-times-shrinking simplicial complex.

\subsection{Relation between arenas and persistence barcodes}
For each raster, the output from \texttt{Dionysus} was a persistence barcode. Each barcode was a collection of subintervals of $[0,5000]$---the persistent \emph{intervals} or \emph{bars}. Among practitioners of persistent homology, the guiding principle is often \cite{carlsson2009topology} the following: short bars correspond to topological noise, and long bars correspond to meaningful topological features. 

At a high level, the number of ``long" bars in a barcode should reflect the number of obstacles in the arena from which the barcode was generated. Some examples are provided in Figure \ref{fig:barcode}. However, there is no ``correct" threshold $\ell$ for counting a bar as long or otherwise; the choice of $\ell$ is dependent on the user. 

For a barcode $B$ and an integer $L \in [0,5000]$, let $\longbars(B,L)$ denote the number of persistent bars in $B$ with length at least $L$. We computed $\longbars(\cdot,4000)$ for each barcode that we obtained from performing the zigzag computation on rasters obtained via the $\binary$ and $\fuzzyb$ models. Plots of this function against $\tau$ are provided in Figure \ref{fig:bar-4000}.

\subsection{1-nearest neighbor classification}
For a fixed choice of firing rate-place field size parameters, we obtained 50 1-dimensional persistence barcodes and computed a $50 \times 50$ bottleneck distance matrix. Next we computed the 1-nearest neighbor classification error rate over 1000 random choices of seed points. The seed points were chosen five at a time, one for each type of arena. After calculating error rates for all parameter pairs, the final result was a plot of error rates vs $\t$, as shown in Figures \ref{fig:error-rates-all} and \ref{fig:error-rates}.

\subsection{Cluster structure via dendrograms}

For each choice of parameter-pair, the 50 resulting 1-dimensional barcodes formed a finite metric space when endowed with the bottleneck distance. We examined the hierarchical clustering structure of this finite metric space by applying single linkage hierarchical clustering and visualizing the result as a dendrogram. Figure \ref{fig:dendro-tau2000} contains the dendrograms for the 15 parameter-pair choices at $\t = 2000$.

\medskip
\paragraph{\textbf{Acknowledgments.}} This work was supported by NSF grant IIS-1422400.

\bibliographystyle{alpha}
\bibliography{biblio}

\newcommand{\etalchar}[1]{$^{#1}$}
\begin{thebibliography}{CGYW14}

\bibitem[BFT{\etalchar{+}}98]{brown1998statistical}
Emery~N Brown, Loren~M Frank, Dengda Tang, Michael~C Quirk, and Matthew~A
  Wilson.
\newblock A statistical paradigm for neural spike train decoding applied to
  position prediction from ensemble firing patterns of rat hippocampal place
  cells.
\newblock {\em The Journal of Neuroscience}, 18(18):7411--7425, 1998.

\bibitem[Bj{\"o}95]{bjorner1995topological}
Anders Bj{\"o}rner.
\newblock Topological methods.
\newblock {\em Handbook of combinatorics}, 2:1819--1872, 1995.

\bibitem[BLIC98]{benke1998modulation}
Tim~A Benke, Andreas Luthi, John~TR Isaac, and Graham~L Collingridge.
\newblock Modulation of ampa receptor unitary conductance by synaptic activity.
\newblock {\em Nature}, 393(6687):793, 1998.

\bibitem[BMD17]{babichev2017robust}
Andrey Babichev, Dmitriy Morozov, and Yuri Dabaghian.
\newblock Robust spatial memory maps encoded in networks with transient
  connections.
\newblock {\em arXiv preprint arXiv:1710.02623}, 2017.

\bibitem[Car09]{carlsson2009topology}
Gunnar Carlsson.
\newblock Topology and data.
\newblock {\em Bulletin of the American Mathematical Society}, 46(2):255--308,
  2009.

\bibitem[CDM17]{cosyne}
Samir Chowdhury, Bowen Dai, and Facundo M{\'e}moli.
\newblock Topology of stimulus space via directed network persistent homology.
\newblock {\em Cosyne Abstracts 2017}, 2017.

\bibitem[CDS10]{zigzag}
Gunnar Carlsson and Vin De~Silva.
\newblock Zigzag persistence.
\newblock {\em Foundations of computational mathematics}, 10(4):367--405, 2010.

\bibitem[CDSM09]{carlsson2009zigzag}
Gunnar Carlsson, Vin De~Silva, and Dmitriy Morozov.
\newblock Zigzag persistent homology and real-valued functions.
\newblock In {\em Proceedings of the twenty-fifth annual symposium on
  Computational geometry}, pages 247--256. ACM, 2009.

\bibitem[CGYW14]{chen2014neural}
Zhe Chen, Stephen~N Gomperts, Jun Yamamoto, and Matthew~A Wilson.
\newblock Neural representation of spatial topology in the rodent hippocampus.
\newblock {\em Neural computation}, 26(1):1--39, 2014.

\bibitem[CI08]{curto2008cell}
Carina Curto and Vladimir Itskov.
\newblock Cell groups reveal structure of stimulus space.
\newblock {\em PLoS Computational Biology}, 4(10), 2008.

\bibitem[CM16]{dowker-arxiv}
Samir Chowdhury and Facundo M{\'e}moli.
\newblock Persistent homology of asymmetric networks: An approach based on
  dowker filtrations.
\newblock {\em arXiv preprint arXiv:1608.05432}, 2016.

\bibitem[Cur17]{curto2017can}
Carina Curto.
\newblock What can topology tell us about the neural code?
\newblock {\em Bulletin of the American Mathematical Society}, 54(1):63--78,
  2017.

\bibitem[DA01]{neuro-book}
Peter Dayan and Laurence~F Abbott.
\newblock {\em Theoretical neuroscience}, volume~10.
\newblock Cambridge, MA: MIT Press, 2001.

\bibitem[DB92]{dudek1992homosynaptic}
Serena~M Dudek and Mark~F Bear.
\newblock Homosynaptic long-term depression in area ca1 of hippocampus and
  effects of n-methyl-d-aspartate receptor blockade.
\newblock {\em Proceedings of the National Academy of Sciences},
  89(10):4363--4367, 1992.

\bibitem[DHL{\etalchar{+}}15]{dong2015long}
Zhifang Dong, Huili Han, Hongjie Li, Yanrui Bai, Wei Wang, Man Tu, Yan Peng,
  Limin Zhou, Wenting He, Xiaobin Wu, et~al.
\newblock Long-term potentiation decay and memory loss are mediated by ampar
  endocytosis.
\newblock {\em The Journal of clinical investigation}, 125(1):234, 2015.

\bibitem[DMFC12]{dabaghian2012topological}
Yu~Dabaghian, Facundo M{\'e}moli, L~Frank, and Gunnar Carlsson.
\newblock A topological paradigm for hippocampal spatial map formation using
  persistent homology.
\newblock {\em PLoS Comput Biol}, 8(8), 2012.

\bibitem[DMS{\etalchar{+}}09]{dabaghian2009topological}
Y~Dabaghian, F~M{\'e}moli, G~Singh, L~Frank, and G~Carlsson.
\newblock Topological stability of the hippocampal spatial map.
\newblock In {\em Front. Syst. Neurosci. Conference Abstract: Computational and
  systems neuroscience}, 2009.

\bibitem[DW91]{dayan1991optimising}
Peter Dayan and David~J Willshaw.
\newblock Optimising synaptic learning rules in linear associative memories.
\newblock {\em Biological cybernetics}, 65(4):253--265, 1991.

\bibitem[EH08]{edelsbrunner2008persistent}
Herbert Edelsbrunner and John Harer.
\newblock Persistent homology-a survey.
\newblock {\em Contemporary mathematics}, 453:257--282, 2008.

\bibitem[EH10]{edelsbrunner2010computational}
Herbert Edelsbrunner and John Harer.
\newblock {\em Computational topology: an introduction}.
\newblock American Mathematical Soc., 2010.

\bibitem[EM14]{edelsbrunner2014persistent}
Herbert Edelsbrunner and Dmitriy Morozov.
\newblock Persistent homology: theory and practice.
\newblock 2014.

\bibitem[FHT01]{friedman-book}
Jerome Friedman, Trevor Hastie, and Robert Tibshirani.
\newblock {\em The elements of statistical learning}, volume~1.
\newblock Springer series in statistics Springer, Berlin, 2001.

\bibitem[Fro92]{frosini1992measuring}
Patrizio Frosini.
\newblock Measuring shapes by size functions.
\newblock In {\em Intelligent Robots and Computer Vision X: Algorithms and
  Techniques}, pages 122--133. International Society for Optics and Photonics,
  1992.

\bibitem[Ghr08]{ghrist2008barcodes}
Robert Ghrist.
\newblock Barcodes: the persistent topology of data.
\newblock {\em Bulletin of the American Mathematical Society}, 45(1):61--75,
  2008.

\bibitem[GPCI15]{giusti2015clique}
Chad Giusti, Eva Pastalkova, Carina Curto, and Vladimir Itskov.
\newblock Clique topology reveals intrinsic geometric structure in neural
  correlations.
\newblock {\em Proceedings of the National Academy of Sciences},
  112(44):13455--13460, 2015.

\bibitem[JKF15]{josselyn2015finding}
Sheena~A Josselyn, Stefan K{\"o}hler, and Paul~W Frankland.
\newblock Finding the engram.
\newblock {\em Nature reviews. Neuroscience}, 16(9):521, 2015.

\bibitem[KMV07]{kemp2007hippocampal}
Anne Kemp and Denise Manahan-Vaughan.
\newblock Hippocampal long-term depression: master or minion in declarative
  memory processes?
\newblock {\em Trends in neurosciences}, 30(3):111--118, 2007.

\bibitem[LDL{\etalchar{+}}16]{liu2016hippocampal}
Yunlong Liu, Shuwen Du, Li~Lv, Bo~Lei, Wei Shi, Yikai Tang, Lianzhang Wang, and
  Yi~Zhong.
\newblock Hippocampal activation of rac1 regulates the forgetting of object
  recognition memory.
\newblock {\em Current Biology}, 26(17):2351--2357, 2016.

\bibitem[LDS89]{lecun1989optimal}
Yann LeCun, John~S Denker, and Sara~A Solla.
\newblock Optimal brain damage.
\newblock In {\em Advances in neural information processing systems}, 1989.

\bibitem[Mor12]{morozov2012dionysus}
Dmitriy Morozov.
\newblock Dionysus.
\newblock {\em Software available at http://www. mrzv. org/software/dionysus},
  2012.

\bibitem[MSB{\etalchar{+}}16]{mcavoy2016modulating}
Kathleen~M McAvoy, Kimberly~N Scobie, Stefan Berger, Craig Russo, Nannan Guo,
  Pakanat Decharatanachart, Hugo Vega-Ramirez, Sam Miake-Lye, Michael Whalen,
  Mark Nelson, et~al.
\newblock Modulating neuronal competition dynamics in the dentate gyrus to
  rejuvenate aging memory circuits.
\newblock {\em Neuron}, 91(6):1356--1373, 2016.

\bibitem[NTB{\etalchar{+}}07]{nicolas2007synapse}
Toni Nicolas, E~Matthew Teng, Eric~A Bushong, James~B Aimone, Chunmei Zhao,
  Antonella Consiglio, Henriette van Praag, Maryann~E Martone, Mark~H Ellisman,
  and Fred~H Gage.
\newblock Synapse formation on neurons born in the adult hippocampus.
\newblock {\em Nature neuroscience}, 10(6):727, 2007.

\bibitem[OD71]{o1971hippocampus}
John O'Keefe and Jonathan Dostrovsky.
\newblock The hippocampus as a spatial map. preliminary evidence from unit
  activity in the freely-moving rat.
\newblock {\em Brain research}, 34(1):171--175, 1971.

\bibitem[PPB{\etalchar{+}}06]{plant2006transient}
Karen Plant, Kenneth~A Pelkey, Zuner~A Bortolotto, Daiju Morita, Akira
  Terashima, Chris~J McBain, Graham~L Collingridge, and John~TR Isaac.
\newblock Transient incorporation of native glur2-lacking ampa receptors during
  hippocampal long-term potentiation.
\newblock {\em Nature neuroscience}, 9(5):602, 2006.

\bibitem[RF17]{richards2017persistence}
Blake~A Richards and Paul~W Frankland.
\newblock The persistence and transience of memory.
\newblock {\em Neuron}, 94(6):1071--1084, 2017.

\bibitem[SDBB15]{spreemann2015using}
Gard Spreemann, Benjamin Dunn, Magnus~Bakke Botnan, and Nils~A Baas.
\newblock Using persistent homology to reveal hidden information in neural
  data.
\newblock {\em arXiv preprint arXiv:1510.06629}, 2015.

\bibitem[SHK{\etalchar{+}}14]{srivastava2014dropout}
Nitish Srivastava, Geoffrey~E Hinton, Alex Krizhevsky, Ilya Sutskever, and
  Ruslan Salakhutdinov.
\newblock Dropout: a simple way to prevent neural networks from overfitting.
\newblock {\em Journal of machine learning research}, 15(1):1929--1958, 2014.

\bibitem[Sin70]{sinai1970dynamical}
Yakov~Grigor'evich Sinai.
\newblock Dynamical systems with elastic reflections.
\newblock {\em Russian Mathematical Surveys}, 25(2):137--189, 1970.

\bibitem[SLH{\etalchar{+}}10]{shuai2010forgetting}
Yichun Shuai, Binyan Lu, Ying Hu, Lianzhang Wang, Kan Sun, and Yi~Zhong.
\newblock Forgetting is regulated through rac activity in drosophila.
\newblock {\em Cell}, 140(4):579--589, 2010.

\bibitem[TLRR15]{tonegawa2015memory}
Susumu Tonegawa, Xu~Liu, Steve Ramirez, and Roger Redondo.
\newblock Memory engram cells have come of age.
\newblock {\em Neuron}, 87(5):918--931, 2015.

\bibitem[TLZ{\etalchar{+}}08]{toni2008neurons}
Nicolas Toni, Diego~A Laplagne, Chunmei Zhao, Gabriela Lombardi, Charles~E
  Ribak, Fred~H Gage, and Alejandro~F Schinder.
\newblock Neurons born in the adult dentate gyrus form functional synapses with
  target cells.
\newblock {\em Nature neuroscience}, 11(8):901--907, 2008.

\end{thebibliography}

\appendix
\section{Topological tools}
\subsection{Persistent homology}
\label{sec:ph}
A good reference for the concepts in this section is \cite{edelsbrunner2010computational}.

Given a finite set $X$, a \emph{simplicial complex} is a collection $K_X$ of nonempty subsets of $X$ such that whenever $\s\in K_X$, any subset $\t\subseteq \s$ also belongs to $K_X$. The singleton elements in this collection are the \emph{vertices} of the simplicial complex, the two-element subsets of $X$ belonging to this collection are the \emph{edges}, and for any $k\in \Z_+$, the $(k+1)$ element subsets of $X$ in this collection are the $k$-simplices. 

Simplicial complexes are useful tools for constructing topological models of discrete data. One can systematically construct simplicial complexes at various resolutions and then study these topological models to understand phenomena in the original dataset at various levels of granularity.  

Given a nested sequence of simplicial complexes $K_1 \subseteq K_2 \subseteq K_3 \subseteq \ldots$, one can perform an operation referred to as \emph{taking homology with field coefficients} to obtain a sequence of finite-dimensional vector spaces with linear maps $H_d(K_1) \r H_d(K_2) \r H_d(K_3) \r \ldots$. The subscript $d$ can be any number $0,1,2,3,\ldots$, and each choice of $d$ leads to a different interpretation of the rank of $H_d(K_\bullet)$. Assuming that the simplicial complexes are constructed from some initial dataset, these interpretations include the number of clusters in the data ($d=0$), the number of loops in the data ($d=1$), and so on.   

The sequence of vector spaces with linear maps above is called a \emph{persistent vector space}. The rank information contained in such an object has a convenient visual representation as a \emph{persistence barcode}, i.e. as a set of lines over a single axis. A 0-dimensional barcode tracks the resolutions at which clusters in data merge together, a 1-dimensional barcode tracks the appearance (and disappearance) of loops in data, and so on. Figure \ref{fig:zz-toy} contains an example of a 1-dimensional barcode.

\subsection{Zigzag persistence}
\label{sec:zigzag}
We refer the reader to \cite{ carlsson2009zigzag, zigzag} for additional details on zigzag persistence.

In standard persistent homology, one of the restrictions on the sequence of simplicial complexes $K_1 \subseteq K_2 \subseteq K_3 \subseteq \ldots$ is that the complexes are only allowed to grow, not reduce. \emph{Zigzag} persistent homology \cite{zigzag} is a generalization of standard persistent homology that overcomes this limitation. In this setting, we have a collection of simplicial complexes $K_1 \leftrightarrow K_2 \leftrightarrow \ldots \leftrightarrow K_n$ where each $\leftrightarrow$ represents either an inclusion $K_{n-1} \subseteq K_n$, or an inclusion $K_{n-1} \supseteq K_n$. Such a sequence is called a \emph{zigzag filtration}. Notice that a zigzag filtration models a dynamic simplicial complex, where simplices are sequentially added or deleted. In particular, the synaptic potentiation complex is an example of a zigzag filtration.
The theory of zigzag persistent homology guarantees that applying homology with field coefficients (in dimension $d$) returns a sequence of vector spaces $\{H_d(K_i)\}_{i=1}^n$ with linear maps in the directions of the corresponding inclusions. A persistence barcode may then be plotted as before. In Figure \ref{fig:zz-toy}, we provide an example of a zigzag filtration and its 1-dimensional barcode. 

All our zigzag persistence computations were carried out using \texttt{Dionysus 1} \cite{morozov2012dionysus}.

\end{document}